\documentclass[epj]{svjour}
\usepackage{graphics}
\begin{document}
%
\title{$J/\psi\to VP$ decays and the quark and gluon content of the\\ $\eta$ and $\eta^\prime$}
\author{Rafel Escribano\inst{1}
}                     
%
%
\institute{Grup de F\'{\i}sica Te\`orica and IFAE, 
Universitat Aut\`onoma de Barcelona, E-08193 Bellaterra (Barcelona), Spain
}
\date{Received: date / Revised version: date}
%
\abstract{
The $\eta$-$\eta^\prime$ pseudoscalar mixing angle and the gluonium content of the
$\eta^\prime$ meson are deduced from an updated phenomenological analysis of $J/\psi$ decays into a vector and a pseudoscalar meson.
In absence of gluonium,
the value of the mixing angle in the quark-flavour basis is found to be $\phi_P=(40.7\pm 2.3)^\circ$.
In presence of gluonium,
the values for the mixing angle and the gluonic content of the $\eta^\prime$ wave function are
$\phi_P=(44.6\pm 4.4)^\circ$ and $Z^2_{\eta^\prime}=0.29^{+0.18}_{-0.26}$, respectively.
The newly reported values of $B(J/\psi\to\rho\pi)$ by the BABAR and BES Collaborations
are crucial to get a consistent description of data.
\PACS{
      {12.39.-x}{Phenomenological quark models}   \and
      {13.25.Gv}{Decays of $J/\psi$}
     } 
} 
\maketitle

\section{Introduction}
\label{intro}
Recently, the KLOE Collaboration reported a new measurement of the
$\eta$-$\eta^\prime$ pseudoscalar mixing angle and the gluonium content of the $\eta^\prime$ meson \cite{Ambrosino:2006gk}.
Combining the value of $R_{\phi}$ with other constraints, they estimated the gluonium content in the $\eta^{\prime}$ wave function as $Z_{\eta^\prime}^2=0.14\pm 0.04$
and the mixing angle in the quark-flavour basis as $\phi_P = (39.7\pm0.7)^{\circ}$.

Later on, we performed a phenomenological analysis of radiative 
$V\to P\gamma$ and $P\to V\gamma$ decays,
with $V=\rho, K^\ast, \omega, \phi$ and $P=\pi, K, \eta, \eta^\prime$,
aimed at determining the gluonic content of the $\eta$ and $\eta^\prime$ wave functions
\cite{Escribano:2007cd}.
We concluded that the current experimental data on $VP\gamma$ transitions indicated within our model a negligible gluonic content for the $\eta$ and $\eta^\prime$.
In particular, accepting the absence of gluonium for the $\eta$, the gluonic content of the 
$\eta^\prime$ wave function amounts to $Z_{\eta^\prime}^2=0.04\pm 0.09$ or
$|\phi_{\eta^\prime G}|=(12\pm 13)^\circ$ and the $\eta$-$\eta^\prime$ mixing angle is found to be
$\phi_P=(41.4\pm 1.3)^\circ$.

Motivated by the discrepancy of the two former analyses,
we perform a new phenomenological analysis of $J/\psi\to VP$ decays to confirm or refute the possible gluonium content of the $\eta^\prime$ meson.
This is now feasible in view of the new experimental data at our disposal which makes of the
$J/\psi\to VP$ decays the most precise and complete set of measurements
(after the $VP\gamma$ transitions) to further test this possibility.
The analogous $\psi^\prime\to VP$ decays will not be considered here since their measurements have larger uncertainties and the set is not complete.

The $J/\psi\to VP$ decays have been studied in the literature (see for instance
Refs.~\cite{Bramon:1997mf,Feldmann:1998vh,Escribano:2002iv,Zhao:2006gw,Li:2007ky,Thomas:2007uy}
and the reviews in Refs.~\cite{Gilman:1987ax,Kopke:1988cs}).
In Ref.~\cite{Bramon:1997mf}, the value of the $\eta$-$\eta^\prime$ mixing angle was deduced from this relevant set of $J/\psi\to VP$ decay data including for the first time corrections due to
non-ideal $\omega$-$\phi$ mixing.
These corrections turned out to be crucial to find $\phi_P=(37.8\pm 1.7)^\circ$, 
which was appreciably less negative than previous results coming from similar analyses.
From the same set of data, a recent determination of the mixing angle assuming no gluonium finds
$\phi_P=(40\pm 2)^\circ$ while with gluonium in the $\eta^\prime$ gives $\phi_P=(45\pm 4)^\circ$ and
$\cos\phi_{\eta^\prime G}=0.84^{+0.10}_{-0.14}$ \cite{Thomas:2007uy}.

\begin{table*}
\begin{center}
\begin{tabular}{cccccc}
\hline
Channel & Exp. & Fit 1 & Fit 2 & Fit 3 & Fit 4 \\[0.5ex]
\hline
$\rho\pi$ 						& $16.9\pm 1.5$		
							& $17.0\pm 1.0$ 			& $16.8\pm 1.1$
							& $17.4\pm 1.0$ 			& $16.9\pm 1.1$ \\[0.5ex]
$\rho^0\pi^0$ 					& $5.6\pm 0.7$		
							& $5.7\pm 0.3$ 			& $5.6\pm 0.4$
							& $5.8\pm 0.3$ 			& $5.6\pm 0.4$ \\[0.5ex]
$K^{\ast +}K^-+\mbox{c.c.}$		& $5.12\pm 0.30$			
							& $5.2\pm 0.4$				& $5.3\pm 0.5$
							& $5.2\pm 0.4$				& $5.2\pm 0.5$ \\[0.5ex]
$K^{\ast 0}\bar K^0+\mbox{c.c.}$ 	& $4.39\pm 0.31$			
							& $4.5\pm 0.5$				& $4.6\pm 0.5$
							& $4.4\pm 0.5$				& $4.5\pm 0.5$ \\[0.5ex]
$\omega\eta$ 					& $1.74\pm 0.20$		
							& $1.59\pm 0.11$			& $1.53\pm 0.24$
							& $1.54\pm 0.11$			& $1.59\pm 0.22$ \\[0.5ex]
$\omega\eta^\prime$ 			& $0.182\pm 0.021$		
							& $0.185\pm 0.060$			& $0.183^{+0.352}_{-0.359}$
							& $0.181\pm 0.060$			& $0.184\pm 0.382$ \\[0.5ex]
$\phi\eta$ 					& $0.75\pm 0.08$		
							& $0.67\pm 0.12$			& $0.68\pm 0.19$
							& $0.73\pm 0.12$			& $0.69\pm 0.17$ \\[0.5ex]
$\phi\eta^\prime$ 				& $0.40\pm 0.07$		
							& $0.38\pm 0.09$			& $0.39^{+0.39}_{-0.41}$
							& $0.39\pm 0.09$			& $0.38^{+0.35}_{-0.36}$ \\[0.5ex]
$\rho\eta$ 					& $0.193\pm 0.023$ 	
							& $0.205\pm 0.021$			& $0.199\pm 0.035$
							& $0.205\pm 0.021$			& $0.196\pm 0.036$ \\[0.5ex]
$\rho\eta^\prime$ 				& $0.105\pm 0.018$ 	
							& $0.117\pm 0.014$			& $0.106^{+0.033}_{-0.043}$
							& $0.116\pm 0.014$			& $0.107^{+0.032}_{-0.043}$
\\[0.5ex]
$\omega\pi^0$ 					& $0.45\pm 0.05$		
							& $0.39\pm 0.03$			& $0.43\pm 0.05$
							& $0.40\pm 0.03$			& $0.44\pm 0.05$ \\[0.5ex]
$\phi\pi^0$ 					& $<0.0064$\ C.L.~90\%	
							& $0.0011\pm 0.0001$		& $0.0012\pm 0.0001$
							& $0$					& $0$ \\[0.5ex]
\hline
\end{tabular}
\caption{
Experimental $J/\psi\to VP$ branching ratios (in units of $10^{-3}$) from
Ref.~\cite{Amsler:2008zz}.
The predicted values with $x=0.81\pm 0.05$ and $\phi_V=(3.2\pm 0.1)^\circ$,
corresponding to Fit 1 (gluonium not allowed) and Fit 2 (gluonium allowed in $\eta^\prime$), and
$x=1$ and $\phi_V=0^\circ$,
corresponding to Fit 3 (gluonium not allowed) and Fit 4 (gluonium allowed in $\eta^\prime$),
are also shown for comparison.}
\label{tableB}
\end{center}
\end{table*}

\section{Notation}
\label{notation}
We work in a basis consisting of the states \cite{Rosner:1982ey}
\begin{equation}
\label{quarkstates}
\begin{array}{l}
|\eta_q\rangle\equiv\frac{1}{\sqrt{2}}|u\bar u+d\bar d\rangle\ ,\qquad
|\eta_s\rangle\equiv |s\bar s\rangle\ ,\\[2ex]
\label{gluoniumstate}
|G\rangle=|\mbox{gluonium}\rangle\ .
\end{array}
\end{equation}
The physical states $\eta$ and $\eta^\prime$ are assumed to be the linear combinations
\begin{equation}
\label{physicaleta}
\begin{array}{rcl}
|\eta\rangle &= &X_\eta |\eta_q\rangle+Y_\eta |\eta_s\rangle+Z_\eta |G\rangle\ ,\\[2ex]
\label{physicaletap}
|\eta^\prime\rangle &=&
X_{\eta^\prime}|\eta_q\rangle+Y_{\eta^\prime}|\eta_s\rangle+Z_{\eta^\prime}|G\rangle\ ,
\end{array}
\end{equation}
with $X_{\eta (\eta^\prime)}^2+Y_{\eta (\eta^\prime)}^2+Z_{\eta (\eta^\prime)}^2=1$.
A significant gluonic admixture in a state is possible only if
$Z_{\eta (\eta^\prime)}^2=1-X_{\eta (\eta^\prime)}^2-Y_{\eta (\eta^\prime)}^2>0$.
The implicit assumptions in Eq.~(\ref{physicaleta}) are the following:
i) no mixing with $\pi^0$ ---isospin symmetry, and
ii) no mixing with radial excitations or $\eta_c$ states.
Assuming the absence of gluonium for the $\eta$,
the coefficients $X_{\eta (\eta^\prime)}$, $Y_{\eta (\eta^\prime)}$ and $Z_{\eta (\eta^\prime)}$
are described in terms of two angles (see Ref.~\cite{Escribano:2007cd} for details),
\begin{equation}
\begin{array}{c}
\label{Xetaetap}
X_\eta=\cos\phi_P\ ,\qquad X_{\eta^\prime}=\sin\phi_P\cos\phi_{\eta^\prime G}\ ,\\[1ex]
\label{Yetaetap}
Y_\eta=-\sin\phi_P\ ,\hspace{1.5em} Y_{\eta^\prime}=\cos\phi_P\cos\phi_{\eta^\prime G}\ ,\\[1ex]
\label{Zetaetap}
Z_\eta=0\ ,\hspace{4.5em}  Z_{\eta^\prime}=-\sin\phi_{\eta^\prime G}\ ,
\end{array}
\end{equation}
where $\phi_P$ is the $\eta$-$\eta^\prime$ mixing angle and
$\phi_{\eta^\prime G}$ weights the amount of gluonium in the $\eta^\prime$ wave-function.
An original approach seeking for phenomenological evidence for the gluon content of
$\eta$ and $\eta^\prime$, based on the effects of $\eta$-$\eta^\prime$ and of axial current anomalies on various decay processes, can be found in Ref.~\cite{Ball:1995zv}.
The standard picture (absence of gluonium) is realized with $\phi_{\eta^\prime G}=0$.
For a comprehensive treatment of $\eta$-$\eta^\prime$ mixing in absence of gluonium
see Refs.~\cite{Feldmann:1998vh,Feldmann:1998sh}.
Some analyses in schemes of mixing between $\eta$, $\eta^\prime$ and glueballs are found in
Refs.~\cite{Rosner:1982ey,Carvalho:1999sw}.
For the vector states $\omega$ and $\phi$ the mixing is given by
\begin{equation}
\label{mixingV}
\begin{array}{rcl}
|\omega\rangle &=& \cos\phi_V|\omega_q\rangle-\sin\phi_V|\phi_s\rangle\ ,\\[2ex]
|\phi\rangle &=& \sin\phi_V|\omega_q\rangle+\cos\phi_V|\phi_s\rangle\ ,
\end{array}
\end{equation}
where $|\omega_q\rangle$ and $|\phi_s\rangle$ are the analog non-strange and strange states of
$|\eta_q\rangle$ and $|\eta_s\rangle$, respectively.

\begin{figure*}
\begin{center}
\resizebox{0.75\textwidth}{!}{%
\centering\includegraphics{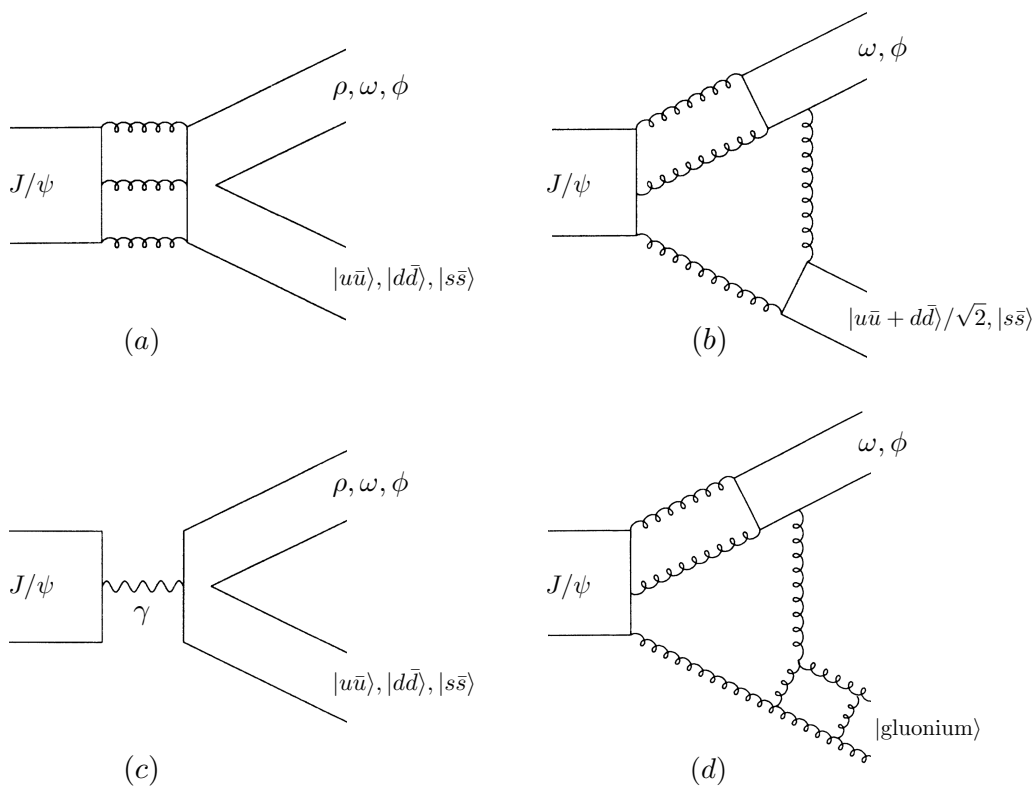}
}  
\caption{Diagrams for $J/\psi\to VP$ decays: (a) singly disconnected strong (SOZI) amplitude;
               (b) doubly disconnected strong (DOZI) amplitude; (c) singly disconnected electromagnetic
               (SOZI EM) amplitude; (d) doubly disconnected production amplitude into a vector and
               a pure glueball state.}
\label{diagrams}
\end{center}
\end{figure*}

\section{Experimental data}
We use the most recent experimental data available for $J/\psi\to VP$ decays taken from
Ref.~\cite{Amsler:2008zz}.
The data for the $\rho\eta$ and $\rho\eta^\prime$ channels remain the same since
1996 \cite{Barnett:1996hr} and were reported by the DM2 \cite{Jousset:1988ni} and
Mark III \cite{Coffman:1988ve} Collaborations.
The new measurements come from the BES Collab.,
Ref.~\cite{Ablikim:2005pr} for $\omega\pi^0$, $\omega\eta$ and $\omega\eta^\prime$ and
Ref.~\cite{Ablikim:2004hz} for $\phi\eta$, $\phi\eta^\prime$ and the upper limit of $\phi\pi^0$,
and the BABAR Collab., Ref.~\cite{Aubert:2007ym} for
$K^{\ast +}K^-+\mbox{c.c.}$ and $K^{\ast 0}\bar K^0+\mbox{c.c.}$,
Ref.~\cite{Aubert:2006jq} for $\omega\eta$ and
Ref.~\cite{Aubert:2007ef} for $\phi\eta$.
The BES data are based on direct $e^+e^-$ measurements,
$e^+e^-\to J/\psi\to\omega\pi^0, \omega\eta, \omega\eta^\prime$ for channels involving $\omega$, 
$e^+e^-\to J/\psi\to\mbox{hadrons}$ for $\phi$, and
the $e^+e^-\to J/\psi\to\phi\gamma\gamma$ for the upper limit of $\phi\pi^0$.
The BABAR data is obtained with the initial state radiation method to lower the center-of-mass energy to the production ($J/\psi$) threshold.
Special attention is devoted to the case of $\rho\pi$.
Four new contributions have been reported since the old weighted average
$B(J/\psi\to\rho\pi)=(1.28\pm 0.10)\%$ \cite{Barnett:1996hr},
$(2.18\pm 0.19)\%$ from $e^+e^-\to\pi^+\pi^-\pi^0\gamma$ \cite{Aubert:2004kj},
$(2.184\pm 0.005\pm 0.201)\%$ from $e^+e^-\to J/\psi\to\pi^+\pi^-\pi^0$ \cite{Bai:2004jn},
$(2.091\pm 0.021\pm 0.116)\%$ from $\psi(2S)\to\pi^+\pi^-J/\psi$ \cite{Bai:2004jn}
---the weighted mean of these two measurements is $(2.10\pm 0.12)\%$, and
$(1.21\pm 0.20)\%$ from $e^+e^-\to\rho\pi$ \cite{Bai:1996rd}.
Thus, the new weighted average is $(1.69\pm 0.15)\%$ with a confidence level smaller than 0.0001 \cite{Amsler:2008zz}.
Finally, the Mark III Coll.~found the ratio $\rho^0\pi^0/\rho\pi$ to be $0.328\pm 0.005\pm 0.027$,
in agreement with the value of one-third expected from isospin symmetry \cite{Coffman:1988ve}.
In Table \ref{tableB}, we show the branching ratios for the different decay channels
according to the present-day values \cite{Amsler:2008zz}.

\section{Phenomenological model}
Since the $J/\psi$ meson is an almost pure $c\bar c$ state, its decays into $V$ and $P$ are 
Okubo-Zweig-Iziuka (OZI) rule-suppressed and proceed through a three-gluon annihilation diagram and an electromagnetic interaction diagram.
Aside from the OZI-suppressed diagrams common to all hadronic $J/\psi$ decays,
the doubly disconnected diagram,
where the vector and the pseudoscalar exchange an extra gluon,
is also expected to contribute to the $J/\psi$ decays, see Fig. \ref{diagrams}.
The doubly disconnected diagram representing the diagram connected to a pure glueball state is also considered.

The amplitudes for the $J/\psi\to VP$ decays are expressed in terms of an
$SU(3)$-symmetric coupling strength $g$ (SOZI amplitude)
which comes from the three-gluon diagram,
an electromagnetic coupling strength $e$ (with phase $\theta_e$ relative to $g$)
which comes from the electromagnetic interaction diagram \cite{Kowalski:1976mc},
an $SU(3)$-symmetric coupling strength which is written by $g$ with suppression factor $r$ contributed from the doubly disconnected diagram (nonet-symmetry-breaking DOZI amplitude) \cite{Haber:1985cv},
and $r^\prime$ which is the relative gluonic production amplitude.
The $SU(3)$ violation is accounted for by a factor $(1-s)$ for every strange quark contributing to
$g$,
a factor $(1-s_p)$ for a strange pseudoscalar contributing to $r$,
a factor $(1-s_v)$ for a strange vector contributing to $r$ \cite{Seiden:1988rr}, and
a factor $(1-s_e)$ for a strange quark contributing to $e$.
The last term arises due to a combined mass/electromagnetic breaking of the flavour-$SU(3)$ symmetry.
This correction was first introduced by Isgur \cite{Isgur:1976hg} who analysed corrections to
$V\to P\gamma$ radiative decays through a parameter $x\equiv\mu_s/\mu_d$ which accounts for the expected difference in the $d$-quark and $s$-quark magnetic moments due to mass breaking.
The $V\to P\gamma$ amplitudes are precisely proportional to the electromagnetic contribution to the $J/\psi\to VP$ decay, whose dominant decay occurs via $J/\psi\to\gamma\to VP$.
These results are reproduced in the $J/\psi\to VP$ amplitudes by means of the identification
$x\equiv 1-s_e$.
The value of the electromagnetic coupling $e$ used here and its counterpart in $V\to P\gamma$ decays are numerically different since in the latter one has a real photon while in the former one considers a virtual photon of mass that of the $J/\psi$.
Other possible electromagnetic contributions such as $J/\psi\to\gamma^\ast P\to VP$ could also be present.
However, as shown in Ref.~\cite{Seiden:1988rr}, these doubly disconnected electromagnetic amplitudes are suppressed by at least and additional factor $\alpha_s$ relative to the singly disconnected electromagnetic amplitude considered here.
The general parameterization of amplitudes for $J/\psi\to VP$ decays is written in Table \ref{tableA}.
In order to obtain the physical amplitudes for processes involving $\omega$ or $\phi$ one has to incorporate corrections due to non-ideal $\omega$-$\phi$ mixing.

\begin{table*}
\begin{center}
\begin{tabular}{cc}
\hline
Process & Amplitude \\[0.5ex]
\hline
$\rho\pi$ 						& $g+e$ \\[0.5ex]
$K^{\ast +}K^-+\mbox{c.c.}$		& $g(1-s)+e(2-x)$ \\[0.5ex]
$K^{\ast 0}\bar K^0+\mbox{c.c.}$ 	& $g(1-s)-e(1+x)$ \\[0.5ex]
$\omega_q\eta$ 				& $(g+e)X_\eta+\sqrt{2}\,r g[\sqrt{2}X_\eta+(1-s_p)Y_\eta]
                                                                                             +\sqrt{2}\,r^\prime g Z_\eta$ \\[0.5ex]
$\omega_q\eta^\prime$ 			& $(g+e)X_{\eta^\prime}
                                                                 +\sqrt{2}\,r g[\sqrt{2}X_{\eta^\prime}+(1-s_p)Y_{\eta^\prime}]
                                                                 +\sqrt{2}\,r^\prime g Z_{\eta^\prime}$ \\[0.5ex]
$\phi_s\eta$ 					& $[g(1-2s)-2e x]Y_\eta
							+r g(1-s_v)[\sqrt{2}X_\eta+(1-s_p)Y_\eta]
                                                                 +r^\prime g (1-s_v)Z_\eta$ \\[0.5ex]
$\phi_s\eta^\prime$ 				& $[g(1-2s)-2e x]Y_{\eta^\prime}
                                                                 +r g(1-s_v)[\sqrt{2}X_{\eta^\prime}+(1-s_p)Y_{\eta^\prime}]
                                                                 +r^\prime g (1-s_v)Z_{\eta^\prime}$ \\[0.5ex]
$\rho\eta$ 					& $3e X_\eta$ \\[0.5ex]
$\rho\eta^\prime$ 				& $3e X_{\eta^\prime}$ \\[0.5ex]
$\omega_q\pi^0$ 				& $3e$ \\[0.5ex]
$\phi_s\pi^0$ 					& $0$ \\[0.5ex]
\hline
\end{tabular}
\caption{
General parameterization of amplitudes for $J/\psi\to VP$ decays.}
\label{tableA}
\end{center}
\end{table*}

Given the large number of parameters to be fitted, 13 in the most general case for 12 observables
(indeed 11 because there is only an upper limit for $\phi\pi^0$),
we initially perform the following simplifications.
First, we set the $SU(3)$-breaking contributions $s_v$ and $s_p$ to zero since they always appear multiplying $r$ and hence the products $r s_v$ and $r s_p$ are considered as second order corrections which are assumed to be negligible.
Second, we fix the parameters $x=m_{u,d}/m_s$ and the vector mixing angle $\phi_V$ to the values obtained from a recent fit to the most precise data on $V\to P\gamma$ decays \cite{Escribano:2007cd},
that is $m_s/m_{u,d}=1.24\pm 0.07$ which implies $x=0.81\pm 0.05$ and $\phi_V=(3.2\pm 0.1)^\circ$.
The value for $x$ is within the range of values used in the literature, $0.62$ \cite{Baltrusaitis:1984rz},
$0.64$ \cite{Jousset:1988ni}, $0.70$ \cite{Morisita:1990cg}, and
the $SU(3)$-symmetry limit $x=1$ \cite{Coffman:1988ve}.
The value for $\phi_V$ is in perfect agreement (magnitude and sign)
with the value $\phi_V=(3.4\pm 0.2)^\circ$ obtained from the ratio
$\Gamma(\phi\to\pi^0\gamma)/\Gamma(\omega\to\pi^0\gamma)$ and the $\omega$-$\phi$ interference in $e^+e^-\to\pi^+\pi^-\pi^0$ data \cite{Dolinsky:1991vq}.
It is also compatible with the value $\phi_V^{\rm quad}=+3.4^\circ$
coming from the squared Gell-Mann--Okubo mass formula (see Ref.~\cite{Amsler:2008zz}).
Finally, we do not allow for gluonium in the $\eta$ wave function,
thus the mixing pattern of $\eta$ and $\eta^\prime$ is given by the mixing angle $\phi_P$ and the coefficient $Z_{\eta^\prime}$.
When possible we will relax these constraints in order to test the goodness of our assumptions.

\section{Results}
We proceed to present the results of the fits.
We will also compare them with others results reported in the literature.
To describe data without considering the contribution from the doubly disconnected diagram
(terms proportional to $rg$) has been shown to be unfeasible \cite{Coffman:1988ve}.
Therefore, it is required to take into account nonet-symmetry-breaking effects.
We have also tested that it is not possible to get a reasonable fit setting the $SU(3)$-breaking correction $s$ to its symmetric value, \textit{i.e.~}$s=0$.
Although it should be easily fixed from the ratio $J/\psi\to K^{\ast +}K^-/K^{\ast 0}\bar K^0$,
the value of $x$ is weakly constrained by the fit
(see Ref.~\cite{Bramon:2000qe} for the case $\phi\to K^+K^-/K^0\bar K^0$).
For those reasons, we start fitting the data with $x=0.81$  (see above) and leave $s$ free.
The vector mixing angle $\phi_V$ is for the time being also set to zero.

\begin{table*}
\begin{center}
\begin{tabular}{ccccc}
\hline
Parameter & Fit 1 & Fit 2 & Fit 3 & Fit 4 \\[0.5ex]
\hline
$g$ &
$1.33\pm 0.04$ & $1.32\pm 0.04$ & $1.35\pm 0.04$ &$ 1.32\pm 0.04$ \\[0.5ex]
$|e|$ &
$0.121\pm 0.004$ & $0.127\pm 0.007$ & $0.121\pm 0.004$ & $0.127\pm 0.007$ \\[0.5ex]
$\arg e$ &
$1.33\pm 0.12$ & $1.34\pm 0.12$ & $1.31\pm 0.12$ & $1.32\pm 0.11$ \\[0.5ex]
$s$ &
$0.28\pm 0.02$ & $0.27\pm 0.03$ & $0.29\pm 0.02$ & $0.27\pm 0.03$ \\[0.5ex]
$r$ &
$-0.38\pm 0.01$ & $-0.43\pm 0.08$ & $-0.37\pm 0.01$ & $-0.36\pm 0.08$ \\[0.5ex]
$\phi_P$ &
$(40.7\pm 2.3)^\circ$ & $(44.6\pm 4.1)^\circ$ & $(40.6\pm 2.3)^\circ$ &  $(45.3\pm 4.1)^\circ$ \\[0.5ex]
$|r^\prime|$ &
--- & $0.04\pm 0.20$ & --- & $0.12\pm 0.22$ \\[0.5ex]
$|\phi_{\eta^\prime G}|$ &
--- & $(32^{+11}_{-22})^\circ$ & --- & $(34^{+10}_{-19})^\circ$ \\[0.5ex]
$Z_{\eta^\prime}^2$ &
--- & $0.29^{+0.18}_{-0.26}$ & --- & $0.31^{+0.17}_{-0.25}$ \\[0.5ex]
$\chi^2/\mbox{d.o.f.}$ &
3.9/5 & 2.6/3 & 3.2/5 & 1.4/3 \\[0.5ex]
\hline
\end{tabular}
\caption{
Results for the fitted parameters with $x=0.81\pm 0.05$ and $\phi_V=(3.2\pm 0.1)^\circ$,
corresponding to Fit 1 (gluonium not allowed) and Fit 2 (gluonium allowed in $\eta^\prime$), and
$x=1$ and $\phi_V=0^\circ$,
corresponding to Fit 3 (gluonium not allowed) and Fit 4 (gluonium allowed in $\eta^\prime$),
are also shown for comparison.}
\label{tableF}
\end{center}
\end{table*}

\subsection{Fits without gluonium}
If gluonium is not allowed in the $\eta^\prime$ wave function,
the result of the fit gives $\phi_P=(40.8\pm 2.3)^\circ$ ---or $\theta_P=(-13.9\pm 2.3)^\circ$---
with $\chi^2/\mbox{d.o.f.}=3.4/5$, in disagreement at the $2\sigma$ level with
$\theta_P=(-19.1\pm 1.4)^\circ$ \cite{Jousset:1988ni},
$\theta_P=(-19.2\pm 1.4)^\circ$ \cite{Coffman:1988ve}, and
$\theta_P\simeq -20^\circ$ \cite{Morisita:1990cg},
but in correspondence with $\phi_P=(39.9\pm 2.9)^\circ$ \cite{Feldmann:1998vh}
and $\phi_P=(40\pm 2)^\circ$ \cite{Thomas:2007uy}.
In some analyses, $x$ is kept fixed to one since it always appears multiplying $e$ and hence also considered as a second order contribution.
In this case, our fit gives $\phi_P=(40.6\pm 2.3)^\circ$ with $\chi^2/\mbox{d.o.f.}=3.2/5$.
However, none of the former analyses include the effects of a vector mixing angle different from zero.
It was already noticed in Ref.~\cite{Bramon:1997mf} that these effects,
which were considered there for the first time, turn out to be crucial to find a less negative value of the $\eta$-$\eta^\prime$ mixing angle.
If we take now the fitted value $\phi_V=+3.2^\circ$ (see above), one gets
$\phi_P=(40.7\pm 2.3)^\circ$ with $\chi^2/\mbox{d.o.f.}=3.9/5$ and
$\phi_P=(40.5\pm 2.3)^\circ$ with $\chi^2/\mbox{d.o.f.}=3.4/5$ for $x=0.81$ and $x=1$, respectively.
These new fits seem to refute the strong correlation between the two mixing angles found in
Ref.~\cite{Bramon:1997mf}.
One interesting feature of the present analysis is the effect produced in the fits by the new averaged value of the $\rho\pi$ branching ratio.
For instance, if $B(\rho\pi)=(16.9\pm 1.5)\%$ \cite{Amsler:2008zz} is replaced by its old value
$(12.8\pm 1.0)\%$ \cite{Barnett:1996hr} one gets $\phi_P=(37.8\pm 1.6)^\circ$
---or $\theta_P=(-16.9\pm 1.6)^\circ$---
with $\chi^2/\mbox{d.o.f.}=8.9/5$ for $x=0.81$ and $\phi_V=3.2^\circ$, 
\textit{i.e.}~the central value and the error of the mixing angle become smaller and the quality of the fit worse.
However, this value is now in agreement with $\theta_P=(-16.9\pm 1.7)^\circ$
found in Ref.~\cite{Bramon:1997mf}.
To further test the effects of the $\rho\pi$ channels we have performed the same fits without them.
The results are
$\phi_P=(40.5\pm 2.6)^\circ$ with $\chi^2/\mbox{d.o.f.}=3.9/3$ for $x=0.81$ and $\phi_V=3.2^\circ$ and $\phi_P=(39.8\pm 2.6)^\circ$ with $\chi^2/\mbox{d.o.f.}=2.8/3$ for $x=1$ and $\phi_V=0^\circ$.
They reveal the preference of the model for the new values of $B(\rho\pi)$.
Concerning the vector mixing angle, if we leave it free then the fitted value is badly constrained and in all cases compatible with zero.
Hence, contrary to the $V\to P\gamma$ case, $J/\psi\to VP$ decays turn out to be inappropriate for fixing this angle.
Finally, we have also tested the effects of the $SU(3)$-breaking contributions $s_p$ and $s_v$.
We get $s_p=-0.08\pm 0.10$ and $s_v=-0.03\pm 0.09$
---with $\phi_P=(39.5\pm 2.6)^\circ$ and $\chi^2/\mbox{d.o.f.}=3.0/3$,
for $x=0.81$ and $\phi_V=3.2^\circ$,
and $s_p=-0.08\pm 0.10$ and $s_v=-0.09\pm 0.10$
---with $\phi_P=(39.9\pm 2.7)^\circ$ and $\chi^2/\mbox{d.o.f.}=2.2/3$,
for $x=1$ and $\phi_V=0^\circ$.
In both cases the effects of these parameters are shown to be negligible thus confirming our hypothesis.
For completeness,
in order to illustrate our results assuming the absence of gluonium in $\eta^\prime$,
we display in Table \ref {tableB} the predicted branching ratios and
in Table \ref{tableF} the fitted values of the model parameters for the cases of 
$x=0.81$ and $\phi_V=3.2^\circ$ (Fit 1) and
$x=1$ and $\phi_V=0^\circ$ (Fit 3), respectively.
As seen in Table \ref{tableF}, the $SU(3)$-violating coupling $s$ amounts to the order of 30\%
while the electromagnetic coupling $e$ to the order of 10\%.
Clearly these contributions play a subdominant role.
Concerning the suppression factor $r$ of the DOZI amplitudes, its value is obviously different from zero and of the order of 40\% in magnitude.
This behaviour corroborates the findings of Ref.~\cite{Coffman:1988ve}.

As stated in the Introduction, we have not performed a similar analysis for $\psi^\prime\to VP$ decays.
However, the $\eta$-$\eta^\prime$ mixing angle should be the same as for the $J/\psi\to VP$ decays.
We can guess the value of this angle by means of the ratio
$\Gamma(\psi^\prime\to\rho\eta^\prime)/\Gamma(\psi^\prime\to\rho\eta)=
(p_{\rho\eta^\prime}/p_{\rho\eta})^3\tan^2\phi_P$.
From Ref.~\cite{Amsler:2008zz}, one gets $\phi_P=(45^{+9}_{-16})^\circ$
which is consistent with the values obtained from $J/\psi\to VP$ decays but with larger uncertainty.

\subsection{Fits with gluonium}
Now, we redo some of the fits accepting a gluonic content in the $\eta^\prime$ wave function.
First, we test this content ignoring the contribution of the gluonic production amplitude,
\textit{i.e.~}$r^\prime=0$.
For $x=0.81$ and $\phi_V=3.2^\circ$
we obtain $\phi_P=(44.6\pm 4.0)^\circ$ and $Z_{\eta^\prime}^2=0.29^{+0.17}_{-0.25}$ with
$\chi^2/\mbox{d.o.f.}=2.6/4$.
Using the mixing angle parameterization (see Notation),
the value obtained for $Z_{\eta^\prime}^2$ is equivalent to
$|\phi_{\eta^\prime G}|=(32^{+10}_{-21})^\circ$.
There is a sign ambiguity in $\phi_{\eta^\prime G}$ that cannot be decided since this angle enters into $X_{\eta^\prime}$ and $Y_{\eta^\prime}$ through a cosine.
If the constraint $Z_\eta=0$ is not imposed then we get $\phi_P=(44.3\pm 4.2)^\circ$,
$|\phi_{\eta G}|=(1.6^{+5.4}_{-6.6})^\circ$ and $|\phi_{\eta^\prime G}|=(34^{+12}_{-21})^\circ$
with $\chi^2/\mbox{d.o.f.}=2.5/3$.
The sign ambiguity is not yet resolved.
However, the only combination allowed by the fit is
$Z_\eta Z_{\eta^\prime}=\sin\phi_{\eta G}\sin\phi_{\eta^\prime G}\cos\phi_{\eta G}<0$.
The fitted value of $\phi_{\eta G}$ supports the hypothesis of disregarding gluonium in the $\eta$
wave-function.
It is worth mentioning the importance of considering asymmetric errors in the minimization procedure to take account of non-linearities in the problem as well as parameter correlations \cite{James:1975dr}.
If not, the values of $Z_{\eta^\prime}^2$ and $\phi_{\eta^\prime G}$ in the case of $Z_\eta=0$
are found to be $0.29\pm 0.20$ and $|\phi_{\eta^\prime G}|=(32\pm 13)^\circ$, respectively.
For $x=1$ and $\phi_V=0$
we obtain $\phi_P=(45.6\pm 4.3)^\circ$ and $Z_{\eta^\prime}^2=0.33^{+0.16}_{-0.26}$
---or $|\phi_{\eta^\prime G}|=(35^{+10}_{-20})^\circ$--- with $\chi^2/\mbox{d.o.f.}=1.8/4$.
If $Z_\eta$ floats then one gets $\phi_P=(45.6\pm 4.2)^\circ$,
$|\phi_{\eta G}|=(1.0^{+8.0}_{-5.6})^\circ$ and $|\phi_{\eta^\prime G}|=(33^{+12}_{-21})^\circ$
(now $Z_\eta Z_{\eta^\prime}>0$) with $\chi^2/\mbox{d.o.f.}=1.7/3$.
As seen, our results are sensitive to the gluonic content in $\eta^\prime$ even in the case of neglecting the contribution of the gluonic production amplitude.
Taking now into account this contribution gives the following results.
For $x=0.81$ and $\phi_V=3.2^\circ$
we find $\phi_P=(44.6\pm 4.1)^\circ$, $Z_{\eta^\prime}^2=0.29^{+0.18}_{-0.26}$
---or $|\phi_{\eta^\prime G}|=(32^{+11}_{-22})^\circ$--- and $|r^\prime|=0.04\pm 0.20$ with
$\chi^2/\mbox{d.o.f.}=2.6/3$.
This time there is a sign ambiguity in $r^\prime$ since it appears in the combination
$r^\prime Z_{\eta^\prime}=-r^\prime\sin\phi_{\eta^\prime G}$ which in this fit is positive.
For $x=1$ and $\phi_V=0$
one gets $\phi_P=(45.3\pm 4.1)^\circ$, $Z_{\eta^\prime}^2=0.31^{+0.17}_{-0.25}$
---or $|\phi_{\eta^\prime G}|=(34^{+10}_{-19})^\circ$--- and $|r^\prime|=0.12\pm 0.22$
(now $r^\prime Z_{\eta^\prime}<0$) with $\chi^2/\mbox{d.o.f.}=1.4/3$.
In summary, these fits seem to favour a substantial gluonic component in $\eta^\prime$ which is, however, compatible with zero at $2\sigma$ due to the large uncertainty.
In all cases, the mixing angle and most of the other parameters are consistent with those assuming no gluonium but with larger uncertainties due to fewer constraints.
The parameter $r^\prime$ weighting the relative gluonic production amplitude is consistent with zero and has a large uncertainty.
These results are in agreement with
the values $\phi_P=(45\pm 4)^\circ$ and $Z^2_{\eta^\prime}=0.30\pm 0.21$ 
---or $|\phi_{\eta^\prime G}|=(33\pm 13)^\circ$---
found in Ref.~\cite{Thomas:2007uy}.
We can also compare with Ref.~\cite{Li:2007ky} when electromagnetic contributions are included.
In their CKM approach a gluonium content in $\eta^\prime$ of 17\% is favoured.
In our scheme this value translates into $\phi_{\eta^\prime G}\simeq 10^\circ$.
The $\eta$-$\eta^\prime$ mixing angle is found to be $\phi_P\simeq 25^\circ$.
In their mixing due to a higher Fock state they get
$\phi_{\eta^\prime G}\simeq 10^\circ$, $\phi_{\eta G}\simeq 2.5^\circ$ and $\phi_P\simeq 31^\circ$.
The values
$\phi_{\eta^\prime G}\simeq 9^\circ$, $\phi_{\eta G}\simeq 9^\circ$ and $\phi_P\simeq 29^\circ$
are obtained in their old perturbation theory scheme.
In all these approaches they find significantly smaller values for both
$\phi_{\eta^\prime G}$ and $\phi_P$.
Finally, we have tested the hypothesis raised in Ref.~\cite{Thomas:2007uy} as to whether a large DOZI amplitude is phenomenologically equivalent to a gluonic component of the $\eta^\prime$.
In this case, for $x=0.81$ and $\phi_V=3.2^\circ$ we obtain an unsatisfactory fit with
$\phi_P=(53.1\pm 1.9)^\circ$, $Z_{\eta^\prime}^2=0.51^{+0.09}_{-0.11}$,
$|r^\prime|=0.81\pm 0.15$, and $\chi^2/\mbox{d.o.f.}=10.5/4$.
For $x=1$ and $\phi_V=0$ we get similar results and $\chi^2/\mbox{d.o.f.}=6.4/4$.
To illustrate our results accepting the presence of gluonium in $\eta^\prime$,
we display in Table \ref {tableB} the predicted branching ratios and
in Table \ref{tableF} the fitted values of the model parameters for the cases of 
$x=0.81$ and $\phi_V=3.2^\circ$ (Fit 2) and
$x=1$ and $\phi_V=0^\circ$ (Fit 4), respectively.

\section{Conclusions}
In summary,
we have performed an updated phenomenological analysis of an accurate and exhaustive set of 
$J/\psi\to VP$ decays with the purpose of determining the quark and gluon content of the
$\eta$ and $\eta^\prime$ mesons.
In absence of gluonium,
we find $\phi_P=(40.7\pm 2.3)^\circ$ for the $\eta$-$\eta^\prime$ mixing angle,
in agreement with recent experimental measurements \cite{Ambrosino:2006gk} and
phenomenological estimates \cite{Thomas:2007uy}.
The inclusion of vector mixing angle effects, not included in previous analyses, turns out to be irrelevant.
We have seen that at present second order contributions are also negligible.
In addition, our fits show a slight preference for the case $x=1$,
which is consistent with the fact of considering this contribution of second order as well.
However, we believe that our main results are the ones obtained for $x=0.81$ and
$\phi_V=3.2^\circ$,
since these values are well understood in $V\to P\gamma$ decays.
Particularly, if $x$ were fixed to 1 then the experimental value of the ratio
$\Gamma(K^{\ast +}\to K^+\gamma)/\Gamma(K^{\ast 0}\to K^0\gamma)$
could not be explained.
We get similar results when the controversial $\rho\pi$ channels are not taken into account.
We have also confirm the significance of incorporating doubly disconnected diagrams in the analysis.
If gluonium is allowed in the $\eta^\prime$ wave function,
we obtain $\phi_P=(44.6\pm 4.1)^\circ$ and $Z^2_{\eta^\prime}=0.29^{+0.18}_{-0.26}$
---or $|\phi_{\eta^\prime G}|=(32^{+11}_{-22})^\circ$,
thus suggesting in this approach a substantial gluonic component in $\eta^\prime$
though not statistically significant at the current experimental precision.
However, if this large gluonic component were confirmed the phenomenological consequences,
for instance in $B\to K\eta^\prime$ decays \cite{Kou:2001pm}, would be relevant.
We have shown that the data is sensitive to this component even in this case of ignoring the effects of a direct gluonic production amplitude.
If these effects are considered then they are seen to be insignificant.
We have verified experimentally that the gluonic content in $\eta$ is negligible.
We have also refute the possibility of a phenomenological equivalence between a large DOZI amplitude and a gluonic component in $\eta^\prime$. 
With respect to the data analysis, we would like to stress once more the importance of considering asymmetric errors in the minimization procedure in order not to draw a somewhat different conclusion.
Finally, we have established that the recent reported values of $B(J/\psi\to\rho\pi)$
by the BABAR and BES Collab.~are crucial to get a consistent description of data.

\begin{acknowledgement}
R.E. gratefully acknowledges the warm hospitality at CERN Theory Division where the analysis was finished.
This work has been supported in part by the Ramon y Cajal program,
the Ministerio de Educaci\'on y Ciencia under grants CICYT-FEDER-FPA2005-02211 and
FPA2008-01430,
the EU Contract No.~MRTN-CT-2006-035482 ``FLAVIAnet'',
the Spanish Con\-solider-Ingenio 2010 Programme CPAN (CSD2007-00042), and
the Generalitat de Catalunya under grant SGR2005-00994.
\end{acknowledgement}

\appendix
\section{Effective description of $J/\psi\to VP$ decays}
\label{appendix}
The effective Lagrangian for the $J/\psi VP$ interaction without the kinematical quantities is expressed as
\begin{equation}
\begin{array}{l}
{\cal L}=\frac{g}{2}\langle\{V,P\}S\rangle+3 e\langle\{V,P\}Q\rangle
+r g\langle V S_v\rangle\langle P S_p\rangle \\[2ex]
\qquad
+r^\prime g\langle V S_v\rangle G\ ,
\end{array}
\end{equation}
where $V$ and $P$ are the matrices containing the vector and pseudoscalar nonets,
\begin{equation}
\begin{array}{l}
V=\left(
\begin{array}{ccc}
\frac{\rho^0}{\sqrt{2}}+\frac{\omega_q}{\sqrt{2}} & 
\rho^+ & K^{\ast +}  \\
\rho^-  & 
-\frac{\rho^0}{\sqrt{2}}+\frac{\omega_q}{\sqrt{2}} & K^{\ast 0}  \\
K^{\ast -}          & 
\bar K^{\ast 0} & \phi_s \\
\end{array}
\right), \\[7ex]
P=\left(
\begin{array}{ccc}
\frac{\pi^0}{\sqrt{2}}+\frac{\eta_q}{\sqrt{2}} & \pi^+                                                                   & K^+ \\
\pi^-                                                                   & -\frac{\pi^0}{\sqrt{2}}+\frac{\eta_q}{\sqrt{2}} & K^0 \\
K^-                                                                     & \bar K^0                                                             & \eta_s \\
\end{array}
\right),
\end{array}
\end{equation}
and $G$ is a pure glueball state.
The matrices $S$, $S_v$ and $S_p$ are introduced to account for the different $SU(3)$ violations,
\begin{equation}
\begin{array}{l}
S=\left(
\begin{array}{ccc}
1 & 0 & 0 \\
0 & 1 & 0 \\
0 & 0 & 1-2s \\
\end{array}
\right),
\quad
S_v=\left(
\begin{array}{ccc}
1 & 0 & 0 \\
0 & 1 & 0 \\
0 & 0 & 1-s_v \\
\end{array}
\right), \\[5ex]
S_p=\left(
\begin{array}{ccc}
1 & 0 & 0 \\
0 & 1 & 0 \\
0 & 0 & 1-s_p \\
\end{array}
\right),
\end{array}
\end{equation}
and $Q=\frac{1}{3}\mbox{diag}(2,-1,-x)$ is the quark charge matrix also modified by a $SU(3)$ violation in the $s$-quark entry.
The general parameterization of amplitudes shown in Table \ref{tableA} are obtained from this Lagrangian after rewriting the mathematical states
$\mbox{\boldmath $\eta$}_{\rm math}^T=(\eta_q,\eta_s,G)$ into the physical states
$\mbox{\boldmath $\eta$}_{\rm phys}^T=(\eta,\eta^\prime,\iota)$ through
$\mbox{\boldmath $\eta$}_{\rm math}^T=\mbox{\boldmath $\eta$}_{\rm phys}^T O$ with
\begin{equation}
O=\left(
\begin{array}{ccc}
X_\eta & Y_\eta & Z_\eta \\
X_{\eta^\prime} & Y_{\eta^\prime} & Z_{\eta^\prime} \\
X_\iota & Y_\iota & Z_\iota \\
\end{array}
\right).
\end{equation}
The various branching ratios, $\mbox{BR(VP)}$,
are simply related to the corresponding amplitude in Table \ref{tableA} by
$\mbox{BR(VP)}=|A\mbox{(VP)}|^2 |\mathbf{p}|^3$,
where $\mathbf{p}$ is the final state recoil momentum in the rest frame of the $J/\psi$.
$\mbox{BR(VP)}$ is given in units of $10^{-3}$ if the momentum is given in GeV.

\end{document}